\documentclass[graybox]{svmult}

\usepackage{mathptmx}       
\usepackage{helvet}         
\usepackage{courier}        
\usepackage{type1cm}        
\usepackage{makeidx}         
\usepackage{graphicx}        
\graphicspath{{Figures/}}
\usepackage{multicol}        
\usepackage[bottom]{footmisc}

\makeindex             

\begin{document}

\title*{Artificial Neural Networks and Fault Injection Attacks$^*$}
\author{Shahin Tajik and Fatemeh Ganji}
\institute{Shahin Tajik and Fatemeh Ganji\at
Worcester Polytechnic Institute, \email{\{stajik,fganji\}@wpi.edu}
\\
\footnotesize{$^*$ This is the authors' version of a survey to appear in the proceedings of the workshop ``AI+Sec: Artificial Intelligence and Security'', which was held 2--6 December, 2019, in Lorentz Center (for more details, see \url{https://www.lorentzcenter.nl/aisec-artificial-intelligence-and-security.html} [Accessed February 11,2021]).} }
%
%
\maketitle

\abstract*{This chapter is on the security assessment of artificial intelligence (AI) and neural network (NN) accelerators in the face of fault injection attacks. 
More specifically, it discusses the assets on these platforms and compares them with those known and well-studied in the field of cryptographic systems. 
This is a crucial step that must be taken in order to define the threat models precisely. 
With respect to that, fault attacks mounted on NNs and AI accelerators are explored.}

\abstract{This chapter is on the security assessment of artificial intelligence (AI) and neural network (NN) accelerators in the face of fault injection attacks. 
More specifically, it discusses the assets on these platforms and compares them with those known and well-studied in the field of cryptography. 
This is a crucial step that must be taken in order to define the threat models precisely. 
With respect to that, fault attacks mounted on NNs and AI accelerators are explored.  
}

\section{Introduction}
Physical attacks can threaten the confidentiality and integrity of the processed data in electronic embedded devices.
Side-channel analysis and fault injection attacks are examples of such physical attacks.
While most side-channel analysis techniques are considered passive, fault injection attacks involve an active adversary.
In other words, the adversary attempts to observe a faulty behavior of the target platform by feeding faulty data or forcing it to operate in a physical condition, which is outside of the specified operating range. 
For instance, by manipulating the supplied voltage (a.k.a., voltage glitching), altering the frequency of the clock signal (a.k.a., clock glitching), or flipping bits in the memory with a laser beam the attacker can cause erroneous operation of the target device~\cite{bar2006sorcerer}.
Fault injection attacks are usually considered a threat if the target platform is in possession of the adversary.
Nevertheless, recent studies~\cite{rowhammer,voltpwn,clkscrew,ramjam,gnad2017voltage} have demonstrated that such fault attacks can also be carried out remotely on different platforms, without any physical access to the victim device.

The primary target of fault injection attacks has been cryptographic devices and secure hardware.
For instance, by injecting faults into a finite state machine (FSM) implemented on a hardware platform, an adversary might be able to bypass some authentication states and get unauthorized access to some security-sensitive states.
Moreover, by injecting faults into the cryptographic operations on a device and using mathematical tools, such as Differential Fault Analysis (DFA), an adversary might be able to recover the secret key by observing the generated faulty ciphertexts.
Cryptographic hardware are not the only targets for 
fault attacks.
The new victims are artificial intelligence (AI)-enabled hardware.

\begin{figure}[t]
	\centering
	\includegraphics[width=0.9\columnwidth]{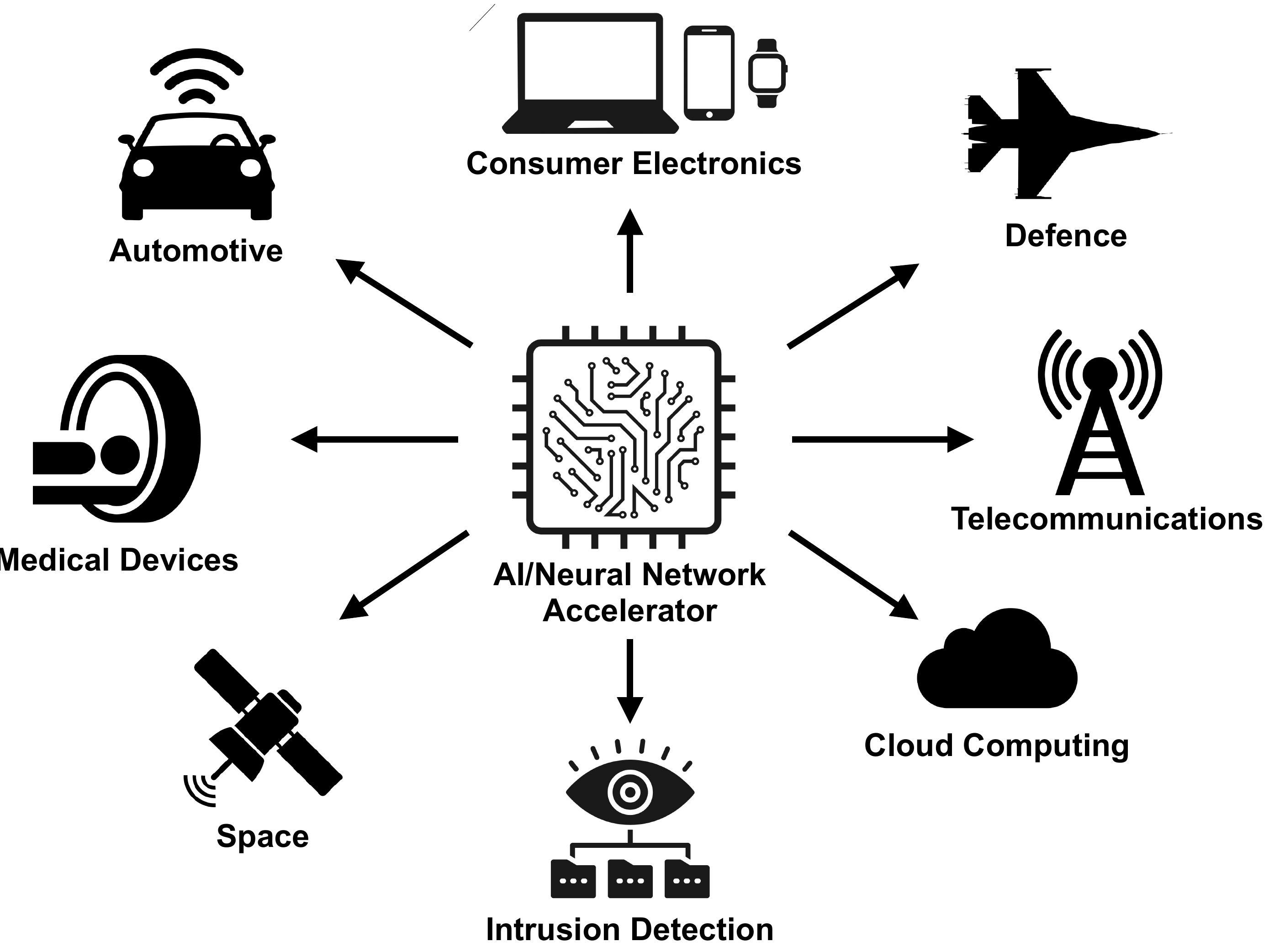}
	\caption{Some applications of AI/neural network accelerators. }
	\label{fig:AIchip}
\end{figure}

Hardware platforms specialized for AI tasks have been applied in various applications and services, offered in the whole spectrum from the cloud to edge, close to the consumer; therefore, such hardware platforms have made commercial, industrial, and defense products feasible. 
End-user devices, such as autonomous cars, smartphones, and robots, are a few examples of such products, see Figure~\ref{fig:AIchip}.
While the benefits of AI in our lives are undeniable, there are several concerns regarding the security of AI hardware.
Similar to cryptographic hardware, AI-enabled hardware could be potential target for physical attacks, such as fault attacks.
The motivations behind physical attacks against the so-called AI/neural network accelerators is manyfold: accessing stored AI assets, IP piracy, obtaining unauthorized access to specific services, or simply disrupting the operation.

Consequently, a great deal of attention has to be paid to protect assets embedded in AI accelerators.
On the positive side, it is known that the artificial neural networks (NNs), which are the most commonly-used AI algorithms implemented on the hardware, are - to some degree - tolerant to faults.
This tolerance to faults might be used as an advantage to replace some of the current fault vulnerable hardware primitives by NNs to obtain a higher level of security against fault attacks, see, e.g.,~\cite{alam2019enhancing}\footnote{Note that fault tolerance is not an intrinsic feature of NNs and should be designed to be exhibited by the models~\cite{survey}.}. 

The aim of this chapter is mainly to review the \emph{vulnerability} of the current AI accelerators and AI IP cores to fault attacks.
The chapter is organized as follows.
In Section~\ref{sec:model}, we review the threat models and assets on AI/neural network accelerators.
In Section~\ref{sec:faultNN}, we discuss the vulnerabilities of NNs to fault attacks.
Section~\ref{sec:accelerators} reviews the available AI/neural network accelerators and their security features.
Potential fault injection techniques against AI/neural network accelerators are presented in Section~\ref{sec:FI_AI}.
Finally, we conclude the chapter in Section~\ref{sec:conclusion}.

\section{Assets and Threat Models}\label{sec:model}
\begin{figure}[t]
	\centering
	\includegraphics[width=0.9\columnwidth]{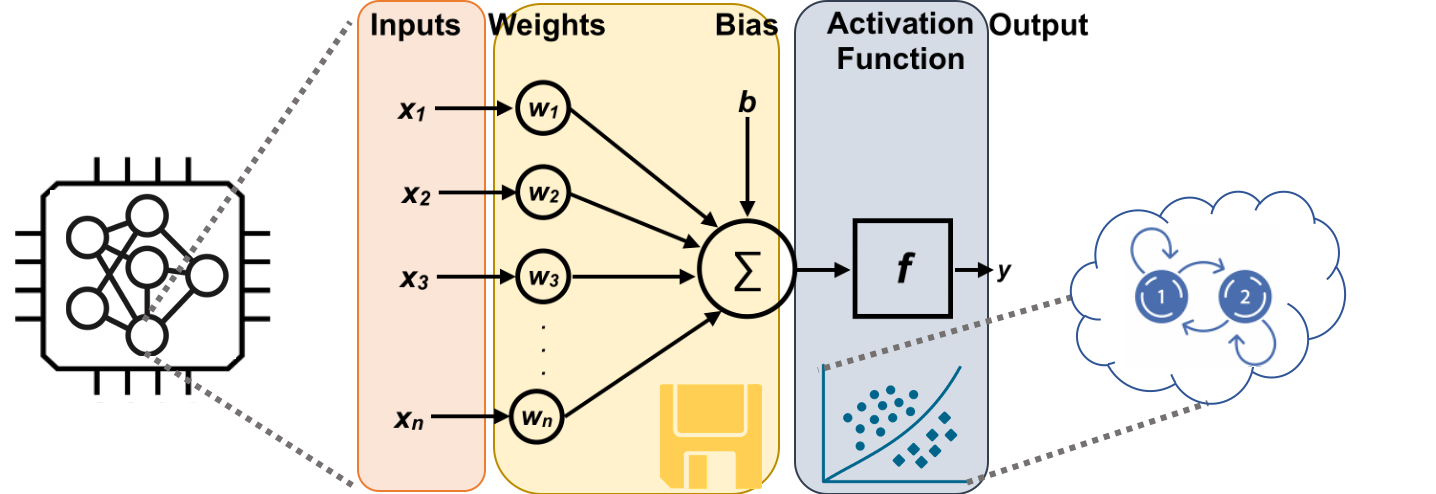}
	\caption {A general architecture of a neural network (NN) , with different abstraction level. The NN is pre-trained and implemented on an FPGA/ASIC. For this, the weights and biases are stored in the memory. The output layer of the NN can be represented by a finite state machine (FSM), cf.~\cite{ramjam}.}
	\label{fig:nn_fsm}
\end{figure}

In order to understand the threat of physical attacks against AI accelerators, we should first understand what the primary assets on AI/neural network accelerators are and how an adversary can benefit from attacking them.
Note that to explore the assets, we should not limit ourselves to any specific class of physical attack (for a more detailed discussion about the impact of fault attack, please see Section~\ref{sec:faultNN}).
Besides, it is helpful to investigate how these assets differ from the conventional assets on cryptographic hardware, which are the traditional targets for fault attacks.
In this regard, we briefly review a few examples of threat models and the main target assets in NNs.
While there exist several learning algorithms that can be realized on hardware, we focus, in this chapter, mainly on artificial NN and its derivatives, since they have gained enormous popularity over the last decade due to their wide range of applications.
An abstract model of NN architecture, which is realized by an FSM in hardware, can be seen in Figure~\ref{fig:nn_fsm}.

\subsection{Attack Scenarios}\label{sec:scenario}
In this subsection, we describe the possible attack scenarios against AI hardware and review the main target assets which have to be protected.

\textbf{Training Data Extraction} 
One of the main assets of an AI system is the training data~\cite{wei2018know}.
In some applications, such training data has to be stored during an enrollment phase on the chip to be compared later in an evaluation phase with the input data.
A prime example of such a scenario is the biometrics data (e.g., fingerprints and face IDs), which are stored on smartphones~\cite{Apple}.
In this case, if an adversary can get access to the training data, she can bypass authentication and get unauthorized access to specific services.
Another example of training data is the medical data gathered from patients, which has to be kept confidential.
Obtaining access to such data might lead to its misuse against the patients.
Even in the cases that the training data is not sensitive, collecting such dataset still costs time and money for the designer of an AI system, e.g., the gathered data for self-driving cars.
Therefore, an adversary can benefit from getting access to such dataset without investing time and money for its collection. 
Finally, the adversary can also tamper with the training data or spoof the training data to change the AI-based system's response. 

\textbf{Structure and Parameters Extraction} 
Another asset is the AI algorithm itself, such as trained neural networks (NNs)~\cite{hua2018reverse,yan2018cache,batina2019csi,sniff}.
For several applications, it is time-consuming to train an NN with large training datasets.
Since such training data is not available to everyone, replicating a similar NN with the same features and classification capabilities might be challenging, and therefore, it needs enormous research and development efforts.
Therefore, to save time and compensate the lack of training data, it can be tempting for an adversary to attack the target AI platform to extract the architecture (e.g., number of layers and neurons) and parameters (e.g., weights, biases, and activation functions) of the trained IP.
In this case, she can clone the IP and produce counterfeits.
For instance, consider AI IP cores, which are used in self-driving cars.
By purchasing one of such cars from a vendor, an adversary (e.g., an end-user or a competitor vendor) gets access to the deployed AI accelerator inside the car with unlimited time to attack and extract the IP and make a profit out of it.
Moreover, in some cases, by knowing the architecture and parameters of an NN, one might be able to recover the used training data. 

\textbf{Disrupting Training and Classification Output}
Many AI platforms are deployed in critical applications, such as intrusion detection systems, autonomous cars, medical devices, and defense systems.
In these cases, any disruption in their performance might lead to irreparable damages or even loss of lives.
It is conceivable that an adversary attempts to mount a physical attack to mislead the AI algorithms during training, inference, or classification phases or merely perform a denial-of-service (DoS) attack~\cite{breier2018practical,ramjam}.

\subsection{AI Assets vs Cryptographic Assets}
As mentioned in Section~\ref{sec:scenario}, there are different types of assets on an AI platform, which need to be protected against physical attacks.
In contrast to AI hardware, cryptographic hardware usually encompasses a single asset, i.e., the secret key, which needs to be protected.
According to \emph{Kerckhoffs’s principle}~\cite{kerckhoffs1883cryptographie,shannon1949communication}, the main assumption is that the security of a device should not rely on the secrecy of its implementation details.
In other words, even if details of a cryptographic implementation are entirely known to the adversary, she should not be able to extract the secret key from the device.
Yet, in the case of an NN, the network structure is itself an asset, and therefore, needs protection.
Besides, for breaking the security of a cryptographic algorithm (i.e., understanding the relation between inputs and outputs), it is needed to extract the exact secret key.
However, an adversary can recover imprecise parameters retrieved from an NN and still generate the same output to a given input.
This diversity of assets in AI hardware makes the security analysis of such systems more complicated than traditional cryptographic hardware. 

\section{Faults in Neural Network}\label{sec:faultNN}
NNs are generally assumed to be tolerant against faults and imprecision to a certain degree~\cite{mahdiani2012relaxed} due to their distributed structures and redundancy.
In other words, NNs can be robust to noisy inputs and demonstrate a low sensitivity to the faults, and therefore, the result of the computation is not drastically affected.
This phenomena is referred to \emph{graceful degradation} in the literature~\cite{survey}.
However, a note of caution is needed here.
Being fault-tolerant highly depends not on only the architecture and size of the NN, but also the training that the NN undergoes. 
For instance, small size NNs might not be fault-tolerant.
Therefore, it cannot be claimed that NNs are in general fault-tolerant~\cite{survey}.

While the threat of fault attacks against NNs is a novel research direction in the field of security, the fault-tolerance feature of the NNs have been studied extensively for years in other areas, including hardware test and machine learning, see, e.g.,~\cite{survey}.  
The main motivation behind these studies in the test community has been the safety assessment of implemented NNs on hardware platforms, which are exposed to transient and permanent faults due to aging, thermal issues, voltage underscaling, and process variations in the very recent nano-scale technologies~\cite{sequin1990fault,bolt1991fault,airforce,mahdiani2012relaxed,salami2018resilience,salami2020experimental,thundervolt}.
These faults can occur in all major parameters of an NN, such as weights, biases, inputs, etc.
On the other hand, in the field of machine learning, several experimental studies have dealt with the concept of adversarial learning, where intentional faults are occurred mainly in the inputs of the NNs to affect the performance of the NN adversely~\cite{goodfellow2014explaining}.
\begin{figure}[t]
	\centering
	\includegraphics[width=0.9\columnwidth]{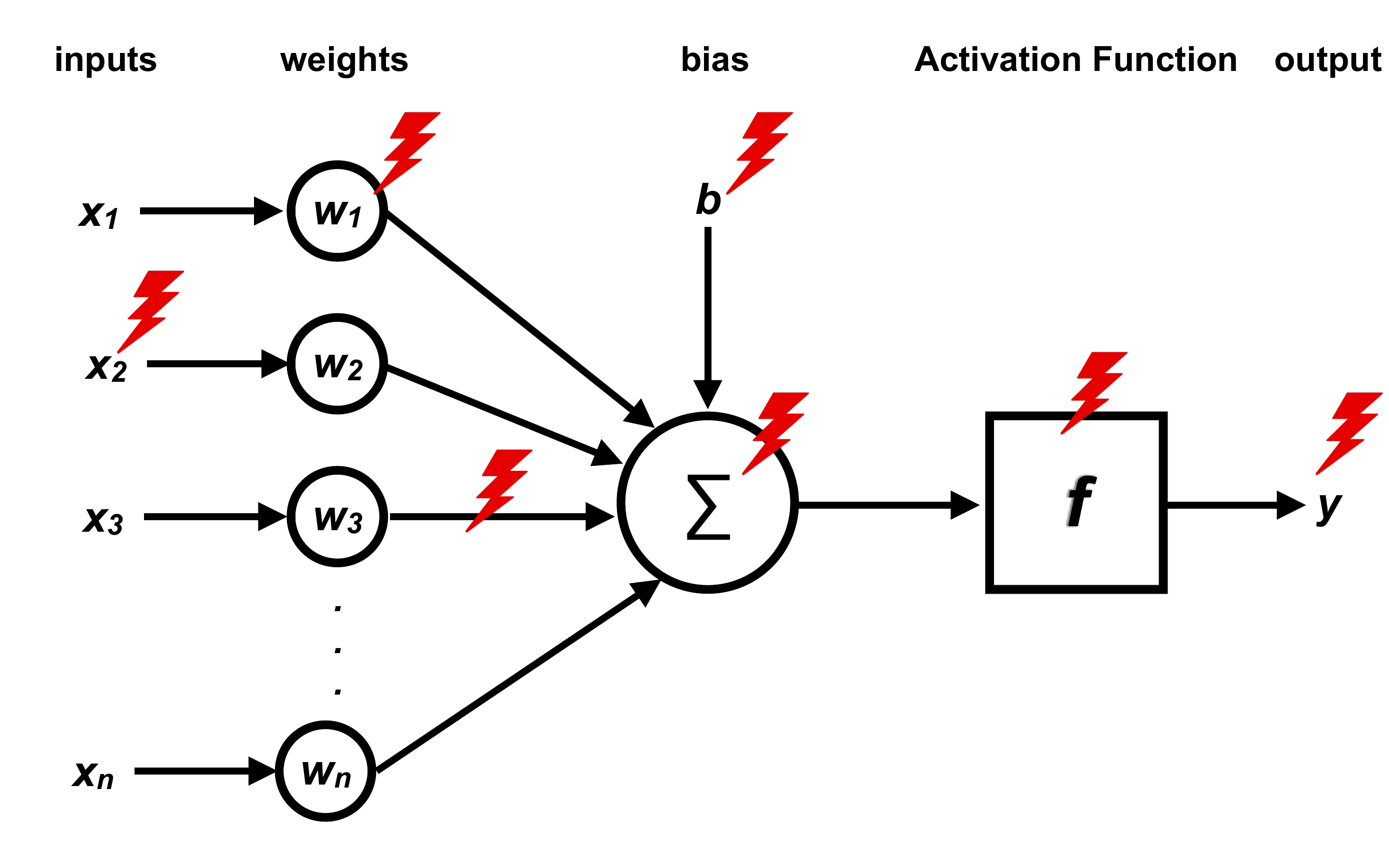}
	\caption{The vulnerable components and parameters of a single neuron to fault injection, inspired by~\cite{survey,sniff}}
	\label{fig:neuron}
\end{figure}

It is indeed possible to inject intentional/adversarial faults in input, hidden, and output layers of an NN implementation (see Figure~\ref{fig:nn_fsm}).
At each layer, the main functional and arithmetic components of each neuron and the parameters involved in the computation can be affected by faults, see Figure~\ref{fig:neuron}.
While manipulating the parameters needs a fault injection into the physical memory of an AI accelerator, e.g., through rowhammer attack~\cite{rowhammer}, affecting multipliers, adders and activation functions, requires usually timing faults, e.g., caused by voltage glitching~\cite{ramjam}.

In this regard, the last two layers of NNs are of great interest to attackers aiming to induce misclassification without perturbing the inputs.
One of the first studies on this has been conducted by Liu et al. to show that by changing some parameters, e.g., the biases, faults can be injected into an NN~\cite{liu2017fault}. 
Since their results were based on simulations, a laser fault injection attack against NNs was introduced in~\cite{breier2018practical} to show to what extent it is possible to obtain similar results in practice.
In their experiments, various functions in an NN have been targeted, and it has been shown that multiple faults can be injected at the same time. 

In another attempt, the attack introduced in~\cite{zhao2019fault} aims to inject faults into trained NNs with two aspects in mind: (1) keeping the classification of other inputs (in this case images), rather than the one targeted by the attacker, unchanged, and (2) minimizing the number of parameter modifications. 
The attack is enhanced by applying a systematic method to determine where a fault should be injected and which parameters (i.e., the biases and weights) can have a stronger influence on the performance of the attack.  
In practice, the issue with determining where the attack should be injected has become more commonly known and discussed in a number of studies. 
As an example,~\cite{rowhammer} has assessed the vulnerability of a large number of NNs to fault attacks and reported the interesting points to be targeted. 
%

This line of research has been pursued by Rakin et al.~\cite{rakin2019bit}, suggesting to inject faults in memories used to store the weights. 
It has been further argued that by applying their proposed algorithm, it could be possible to identify the most vulnerable bits of weights saved in the memory to achieve a high-performance attack. 
For this purpose, and to verify that in practice, trained NNs have been employed to conduct simulations. 
Moreover, the attack has launched in a white-box fashion, i.e., full access to the weights and gradients is required. 
The same authors have shown that the above attack can be combined by an adversarial attack to launch a Trojan attack against NNs~\cite{rakin2020tbt}. 
In doing so, when a trigger is enabled (i.e., giving a specially designed input) the network is forced to classify all inputs to a certain target class.
It is expected that when following this procedure, an adversarial attack could become more efficient as solely a set of randomly sampled data is required to prepare the trigger. 
Additionally, access to neither the training data, nor the training method and its hyperparameters used during training is required. 
This has been demonstrated by performing simulation on NNs, similar to~\cite{rakin2019bit}. 

%

The key message of this section is that the attacks discussed above have been launched to either target an accessible trained model through simulation or inject faults into hardware made available to the attacker.
Next, Section~\ref{sec:FI_AI} reviews attacks in a more realistic scenario, i.e., remotely. 
Such attacks usually targeted AI accelerators; hence, before elaborating on the attacks against them, it is necessary to explain the architectures and IP protection mechanisms associated with accelerators as addressed in the next section. 

\section{AI/Neural Network Accelerators}\label{sec:accelerators}
The primary features of AI accelerators, which differentiate them from the conventional microprocessors, are the capability to perform parallel processing of data and low-precision arithmetic.
Examples of conventional devices, which can fulfill these requirements include graphics processing units (GPUs) and field programmable gate arrays (FPGAs).
Moreover, novel neural network processors are emerging to accelerate AI tasks on the edge devices.
To assess the security of the AI IP cores and stored assets on these platforms, below, we briefly review their security.

\subsection{GPUs}
GPUs are designed primarily to run intensive computational tasks for image processing, which require parallelism. 
Since these platforms are offering parallel computing via several cores, they have been adapted for training and inference tasks in AI applications as well.
GPUs are usually controlled by general purpose processors, such as  x86  and  ARM  processors, in the same chip package.
In this case, the security of stored AI IP cores and data is reduced to the security of the entire processor platform.
An adversary, interested in extracting the NN structure or its parameters inside the memory, could attack the firmware protection schemes to obtain access to the IP. 
On the other hand, security features, such as ARM TrustZone or Intel SGX, are utilized to assure the confidentiality and integrity of the data in the memory of the chip during runtime.
Naturally, the attacker can still mount physical attacks during the runtime on these platforms to interfere with the AI algorithm computation.

\subsection{FPGAs}
FPGAs provide the high parallelism of hardware with the reconfiguration flexibility of the software.
FPGAs contain several thousands of configurable logic resources, a large number of dedicated digital signal processors, and dual-port RAMs.
Hence, they are suitable for the realization of large parallel neural networks.
Most FPGAs do not have any non-volatile memory (NVM) inside their packages to store their configuration data, the so-called, i.e., the so-called bitstream.
Therefore, the bitstream has to be stored in an external NVM and transferred into the FPGA upon each power-on.
The AI IP cores are also included in the bitstream.
Hence, the confidentiality and integrity of AI IP cores and its parameters depend on the security of the bitstream encryption and authentication schemes.
Moreover, an attacker can still tamper with the FPGA during runtime and mount physical attacks, such as fault attacks for reconfiguration of device, to influence the behavior of the AI algorithm. 

\subsection{Custom AI/Neural Network Accelerators}
Instead of using a general-purpose processor or programmable logic, several research teams and vendors have started to design and fabricate their own AI/neural network/deep learning accelerators.
DianNao~\cite{diannao}, EIE~\cite{eie}, Eyeriss~\cite{eyeriss}, Prime~\cite{prime}, and MAERI~\cite{maeri} are a few examples of such designed neural network accelerators by academia.
On the other hand, products such as Google Tensor Product Unit (TPU)~\cite{tpu}, Tesla Hardware 3 platform~\cite{hw3}, and Apple Neural Engine~\cite{applenn} are examples of proprietary neural network accelerators designed by vendors.
These co-processors contain dedicated neural network processors, which are specifically optimized for AI tasks.
For instance, modules, such as activation functions, multiplication and addition operations, are realized in hardware to speed up the computation.
These engines are integrated internally or externally to general-purpose processors based on ARM or x86 architectures.
In this case, since the AI IP is included in the firmware, the confidentiality and integrity of the stored data and model is reduced to the security of the firmware protection schemes on these platforms.
Since these accelerators are usually using compatible silicon-based digital logic and memory technologies, their vulnerability to fault attacks during the runtime is similar to other conventional platforms.

\section{Fault Injection Attacks on AI Accelerators}\label{sec:FI_AI}
As the threat of (passive) side-channel analysis, such as power analysis and cache attacks, against AI accelerators has been discussed extensively in the literature~\cite{batina2019csi,yan2018cache,wei2018know,hua2018reverse,dubey2019maskednet}, here we focus on vulnerabilities of AI algorithms and accelerators to fault attacks.

In order to understand how an adversary can mount fault injection attack against an AI accelerator, we consider two different scenarios: (i) the adversary has physical access to the target, and hence, she can feed arbitrary inputs or operate the target in a non-valid condition (ii) the target is accessible only remotely, and the adversary can only interact with it by sending inputs and run her own code in a virtualized environment. 

\subsection{Traditional Fault Attack}
In the case of traditional fault attacks, it is assumed that the adversary possesses the target.
The neural network accelerators in the edge devices (e.g., smartphones, autonomous vehicles, or any other IoT device) are giving such access to end-users.
Thus, based on the adversary's capability, different classes of fault attacks can be mounted on AI-enabled devices.
The least expensive fault attacks are non-invasive fault injection attacks, e.g., voltage glitching~\cite{salami2018resilience,salami2020experimental}, clock glitching~\cite{liu2020imperceptible,liu2020stealthy}, and electromagnetic faults, which can cause timing faults into the control and data path of the AI accelerator, and therefore, cause faulty computations or even erroneous outputs.
More sophisticated fault attacks are based on more invasive techniques.
The most prominent technique in this class is the laser fault injection attacks, which enables an adversary to not only influence the timing of the signal on the chip but precisely flip bits inside the memories, such as SRAMs.
In this case, the attacker has potentially a higher control or manipulation capability over the AI co-processor's behavior~\cite{hou2020security}.

\subsection{Remote Fault Attacks}
\begin{figure}[t]
	\centering
	\includegraphics[width=0.9\columnwidth]{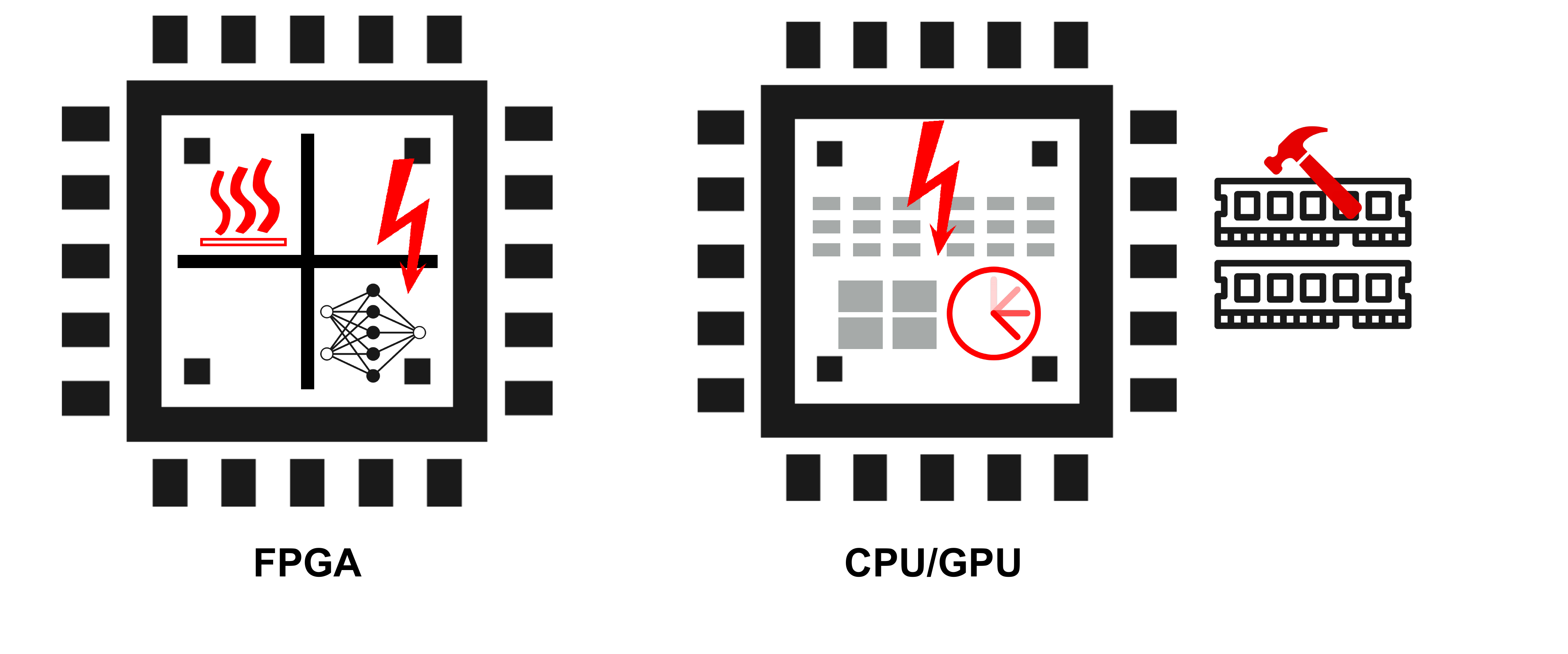}
	\caption{Remote Fault Attacks on AI accelerators: Trojans inside an IP core (e.g., RAM-Jam and power waster circuits) can cause overheating and voltage drop inside FPGAs and cause timing faults into neural networks realized in adjacent IP cores. On processors, remote attacks, such as rowhammer can inject faults into DRAM modules and manipulates the NN parameters. Other attacks, e.g., clkscrew and v0ltpwn attacks, can cause timing faults into the NN computation.}
	\label{fig:remote}
\end{figure}
The assumption that the adversary has physical access to the target chip might be valid for edge devices.
However, in other scenarios, where the chip is deployed, for instance, in the cloud, the adversary cannot perform the traditional fault injection attacks on the target.
In this scenario, the adversary is usually able to rent shared resources (i.e., processing power and memory) on the cloud chips.
Several studies have demonstrated that even with this limited access to the target chip, the adversary can perform fault injection attacks, see Figure~\ref{fig:remote}.

In the case of FPGAs, the resources of the chip can be spatially shared between several IP cores.
The adversary can get partial access to a multi-tenant FPGA in a cloud to implant a hardware Trojan. 
The adversary's goal here is to remotely activate the hardware Trojan and inject a fault from her IP into other adjacent IP cores.
The IP cores on multi-tenant FPGAs are separated by moats, and hence, there is no direct communication link between them. 
However, the Power Distribution Network (PDN) of the FPGA is shared between these IP cores. 
In order to bypass the isolation between IP cores and inject faults into neighboring IPs (e.g., an NN), the adversary has to load valid designs, which contain stealthy hardware Trojans capable of dropping the core voltage and overheating the chip, see Figure~\ref{fig:remote}.
To avoid the detection carried out by bitstream security checks, the hardware Trojan has to be implanted in a valid configuration (i.e., a bitstream).
RAM-Jam~\cite{ramjam} is an example of such Trojans, which causes memory collisions in the dual-port RAMs of the FPGA and creates transient short circuits, leading to voltage drop and temperature increase on the chip. 
As a result, as one of the first attacks launched remotely\footnote{In an independent study conducted in parallel with the one introducing RAM-Jam, Hong et al. have proposed their attack~\cite{rowhammer}.}, RAM-Jam can inject faults into pre-trained NN engines. 
For this purpose, the attack has targeted the hidden and the output layers of an NN by flipping bits (i.e., weights of NN) and bypassing the transitions between states representing the output classes. 

Similarly, power waster circuits based on Ring-Oscillators and other valid power-hungry circuits~\cite{gnad2017voltage,provelengios2019characterizing,powerwaster} can cause severe voltage drops inside FPGAs.
In both cases, the resulting voltage and temperature changes can cause timing faults in the neural network computations, running in parallel on the same FPGA, but owned by another user.

Similar to multi-tenant FPGAs, the deployed CPUs or GPUs as AI accelerators in the cloud are vulnerable to remote fault attacks as well.
For instance, rowhammer attack~\cite{rowhammer} enables the attacker to flip bits in DRAM connected to the AI accelerators, responsible for storing the neural network parameters.
Moreover, attacks, such as v0ltpwn~\cite{voltpwn} and clkscrew~\cite{clkscrew} can alter the voltage and clock frequency of the processor, respectively.
As a result, timing faults can be injected into the operation of the neural network calculations, running in another virtual machine on the same processor.
Note that most of these attacks are demonstrated on CPUs and have not been tested on GPUs yet.
One reason could be the differences in the schedulers of GPUs, which decide on the resource allocation in real time.
The uncertainty resulting from this real-time memory and computing core allocation makes fault attacks more challenging or ineffective.
These schedulers are proprietary, and even though there are some reverse-engineering studies, the knowledge on how they work is limited.

\section{Conclusion}\label{sec:conclusion}
This chapter reviewed the threat of physical attacks, especially fault injection attacks, against neural networks and their implementations on AI accelerators.
We explored the primary assets of an AI system that needs to be protected against physical attacks.
Moreover, we investigated the potential attack scenarios against an AI system both from theoretical and practical points of view, and how an adversary can benefit from such attacks.

Besides, we discussed and compared the assets on an AI accelerator with those on conventional cryptographic hardware.
It became evident that the AI assets are more diverse, and therefore, the formal security assessment of such systems is more complicated.
Consequently, further research on the impact of fault attacks on AI systems is needed to identify potential vulnerabilities and design proper countermeasures against them.
Finally, since neural networks demonstrate some fault resiliency in specific scenarios, it is interesting to further evaluate the applicability and practicality of replacing fault vulnerable hardware/software primitives by a more fault-tolerant neural network.

\bibliographystyle{spmpsci}
\bibliography{reference}
\end{document}